%% file: main.tex
\documentclass[sigconf]{acmart}
\renewcommand\footnotetextcopyrightpermission[1]{} 
\usepackage{textcomp}
\usepackage{graphicx}
\usepackage{xspace}
\usepackage{tikz}
\usepackage{cprotect}
\usepackage{algorithmicx}
\usepackage[plain]{algorithm}
\usepackage{algpseudocode}
\usepackage{listings}
\usepackage{multirow}
\usepackage{pifont}
\usepackage{enumitem}
\usepackage{subcaption}
\usepackage{mathtools}
\usepackage{amsmath}
\usepackage{scalerel}
\usepackage{wrapfig}
\usepackage{array}
\usepackage{adjustbox}
\usepackage{wasysym}
\usepackage{dblfloatfix}

\AtBeginDocument{%
  }

\setcopyright{none}

\begin{document}

\title{A Unified Architecture for Efficient Binary and Worst-Case Optimal Join Processing}

\author{Amirali Kaboli}
\email{amirali.kaboli@ed.ac.uk}
\affiliation{
  \institution{University of Edinburgh}
  \city{Edinburgh}
  \country{United Kingdom}
}

\author{Alex Mascolo}
\email{a.mascolo@ed.ac.uk}
\affiliation{
  \institution{University of Edinburgh}
  \city{Edinburgh}
  \country{United Kingdom}
}

\author{Amir Shaikhha}
\email{amir.shaikhha@ed.ac.uk}
\affiliation{
  \institution{University of Edinburgh}
  \city{Edinburgh}
  \country{United Kingdom}
}

\input{macros}

\begin{abstract}
Join processing is a fundamental operation in database management systems; however, traditional join algorithms often encounter efficiency challenges when dealing with complex queries that produce intermediate results much larger than the final query output.
The emergence of worst-case optimal join (WCOJ) algorithms represents a significant advancement, offering asymptotically better performance by avoiding the enumeration of potentially exploding intermediate results.
In this paper, we propose a unified architecture that efficiently supports both traditional binary joins and WCOJ processing.
As opposed to the state-of-the-art, which only focuses on either hash-based or sort-based join implementations, our system accommodates both physical implementations of binary joins and WCOJ algorithms.
Experimental evaluations demonstrate that our system achieves performance gains of up to $3.1\times$ (on average $1.5\times$) and $4.8\times$ (on average $1.4\times$) over the state-of-the-art implementation of Generic Join and Free Join methods, respectively, across acyclic and cyclic queries in standard query benchmarks.
\end{abstract}

\input{acm_blocks}

\maketitle

\input{sections/1_introduction}
\input{sections/2_background}
\input{sections/3_system}
\input{sections/4_efficiency}

\input{sections/5_experiments}

\input{sections/6_conclusion}


\section*{Acknowledgements}
The authors thank Huawei for their support of the distributed
data management and processing laboratory at the University of Edinburgh. This work was funded in part by a gift from RelationalAI.

\bibliographystyle{ACM-Reference-Format}
\bibliography{refs}

\end{document}

%% file: macros.tex
\newcommand{\smartpara}[1]{\noindent \textbf{#1.}}
\newcommand{\myrightarrow}{$\rightarrow$\xspace}
\newcommand{\myleadsto}{$\leadsto$\xspace}

\newcommand{\todo}[1]{{\color{RedOrange} TODO: #1 }}
\newcommand{\alex}[2]{{\color{OliveGreen} Alex: #2 }}

\newcommand{\sdql}{SDQL\xspace}
\newcommand{\wcoj}{WCOJ\xspace}
\newcommand{\cpp}{C++\xspace}
\newcommand{\fj}{Free Join\xspace}
\newcommand{\gj}{Generic Join\xspace}
\newcommand{\job}{Generic Join\xspace}
\newcommand{\clover}{$Q_{\clubsuit}$\xspace}
\newcommand{\figref}[1]{Figure~\ref{fig:#1}\xspace}
\newcommand{\figsref}[2]{Figures~\ref{fig:#1} and~\ref{fig:#2}\xspace}
\newcommand{\secref}[1]{Section~\ref{sec:#1}\xspace}
\newcommand{\secsref}[2]{Sections~\ref{sec:#1} and~\ref{sec:#2}\xspace}
\newcommand{\equref}[1]{Eq (\ref{eq:#1})\xspace}
\newcommand{\relx}[2]{$\sigma_{x=#2}(#1)$}
\newcommand{\smallvec}{SmallVector\xspace}
\newcommand{\atom}[4]{$#1_#2(#3_#4)$\xspace}
\newcommand{\tab}{\;\;\;}
\newcommand{\delim}{\tab $\mid$ \tab}

\colorlet{myblue}{blue!70!black}
\colorlet{comgreen}{green!10!gray}
\colorlet{mygreen}{green!40!black}
\definecolor{added}{rgb}{0.8,1,0.8}
\definecolor{deleted}{rgb}{1,0.8,0.8}
\newcommand{\codesize}{\small}
\newcommand{\codestyle}{\codesize\ttfamily\color{black}}
\newcommand{\commentstyle}{\codesize\ttfamily\color{comgreen}}
\newcommand{\defines}{\triangleq}
\newcommand{\codekwstyle}{\codesize\ttfamily\bfseries\color{myblue}}
\newcommand{\codemacrostyle}{\codesize\ttfamily\bfseries\color{mygreen}}
\newcommand{\code}[1]{\texttt{#1}}
\newcommand{\codekw}[1]{\textrm{\codekwstyle{}#1}}
\newcommand{\codemacro}[1]{\textrm{\codemacrostyle{}#1}}
\newcommand{\zeromacro}[1]{\codemacro{zero}\code{[}#1\code{]}}
\newcommand{\grammarcomment}[1]{\textit{\small #1}}

\lstdefinelanguage{sdql}{
    morekeywords={
        if,then,else,let,in,not,sum,range,
        int,bool,promote,true,false,
        @vec,@smallvec,@vecdict,@smallvecdict,@st,@range
    },
    basicstyle=\codestyle,
    keywordstyle=\codekwstyle,
    commentstyle=\commentstyle,
    numbers=left,
    numbersep=2pt,
    numberstyle=\tiny\color{gray},
    sensitive,
    morecomment=[l]//,
    breaklines=true,
    mathescape=true,
    breakatwhitespace=true,
    xleftmargin=1em,
    xrightmargin=3em,
    aboveskip=0pt,
    belowskip=0pt,
    lineskip=0.5pt,
    columns=fullflexible,
    keepspaces=true,
}[keywords,comments,strings]
\lstset{language=sdql}
\lstMakeShortInline[columns=fixed, keepspaces=true, language=sdql]!

\lstset{
    language=C++,
    keywordstyle=\codekwstyle,
    columns=fullflexible,
    keepspaces=true,
    commentstyle=\commentstyle,
    firstnumber=1,
    numbers=left,
    numbersep=2pt,
    numberstyle=\tiny\color{gray},
    xleftmargin=1em,
    xrightmargin=3em,
}

%% file: acm_blocks.tex


\keywords{Worst-case optimal join, query compilation, generic join, free join, hashing versus sorting.}


%% file: sections/1_introduction.tex
\section{Introduction}\label{sec:introduction}

Efficient query processing is essential for the performance of modern database management systems, particularly as the complexity and size of data continue to grow.
Join operations, which combine records from multiple tables, play a pivotal role in this process; however, traditional join algorithms often face significant efficiency challenges when processing complex queries that produce intermediate results much larger than the final output.
The emergence of worst-case optimal join (\wcoj) algorithms~\cite{wcojalg1, wcojalg2, leapfrog, skew} represents a significant advancement, offering asymptotically better performance by avoiding the enumeration of potentially exploding intermediate results.
It can be the case for both cyclic and acyclic queries as opposed to the common belief that \wcoj is designed for cyclic queries.

There have been previous efforts~\cite{umbra, emptyheaded, graphflow} to adopt an approach where \wcoj algorithms are used for certain parts of a query while traditional join algorithms are applied to the rest.
However, the use of two distinct algorithms within the same system introduces additional complexity, which has hindered the broader adoption of \wcoj methods.
The current state-of-the-art system, and our main competitor in this work, is \fj~\cite{freejoin}, which aims to address this challenge by unifying \wcoj with traditional join methods.
\fj provides a platform capable of executing a wide range of join queries, offering performance benefits in both \wcoj and traditional join scenarios.
However, despite these advances, \fj supports only a specific class of \wcoj algorithms--hash-based approaches--limiting its coverage and flexibility in handling other algorithmic paradigms such as sort-based joins.

Our approach stands out by leveraging advancements in programming languages, specifically through the use of \sdql~\cite{sdql}, an intermediate language designed for functional collection programming using semi-ring dictionaries.
By employing \sdql as our intermediate representation, we can translate worst-case optimal join (\wcoj) queries into an intermediate representation that can be directly compiled into highly efficient \cpp code.
This gives our system a key advantage in flexibility and performance optimization.

The use of \sdql enables us to introduce a unified architecture that efficiently supports both traditional binary joins and \wcoj algorithms.
Moreover, our system can handle hash-based and sort-based paradigms of \wcoj processing, significantly improving over state-of-the-art systems such as \fj.
Existing systems typically only focus on one approach -- either hash-based or sort-based -- while we provide support for both, ensuring that our system can adapt to various input data types and query execution scenarios.
This broad capability enhances the versatility and overall performance of our system across a wide range of join queries.
We demonstrate that our system not only matches but also outperforms the performance of the \fj framework as the state-of-the-art.

In this paper, we present the following contributions:
\begin{enumerate}
    \item A unified architecture that integrates both efficient binary join and WCOJ processing (\secref{system}).
    \item A comprehensive set of optimizations that refine our initial implementation into a highly optimized system (\secsref{opts:dictspec}{opts:earlyproj}).
    \item A novel hybrid approach, along with support for sort-based WCOJ algorithms, leverages the strengths of both hash-based and sort-based paradigms (\secref{opts:sorting}).
    \item A detailed experimental evaluation of our system and the applied optimizations (\secref{exp}). Specifically,  we show that our method achieves speedups of up to $3.1\times$ and $4.8\times$ compared to the \gj and \fj implementations within the \fj framework, respectively.
\end{enumerate}


%% file: sections/2_background.tex
\section{Background}\label{sec:background}

A full conjunctive query is expressed in the form shown in \equref{conjq}.
In this notation, each term \atom{R}{i}{x}{i} is referred to as an atom, where $R_i$ represents a relation name, and $x_i$ is a tuple of variables.
The query is considered full because the head variables $x$ encompass all variables that appear in the atoms.
We assume that selections have been pushed down to the base tables, meaning that each atom $R_i$ may already incorporate a selection over a base relation.
Similarly, projections and aggregations are performed only after the full join operation is executed, and hence, they are not explicitly included in \equref{conjq}.
\begin{equation}
Q(x) :- R_1(x_1), \dotsc, R_m(x_m)
\label{eq:conjq}
\end{equation}

\smartpara{Example}
Throughout this paper, we will make use of a conjunctive query referred to as \clover, denoted in \equref{clover}.
\begin{equation}
Q_{\clubsuit}(x, a, b) :- R(x, a), S(x, b), T(x)
\label{eq:clover}
\end{equation}
The SQL query of \clover is as below and its corresponding implementation using binary joins in \sdql and \cpp is shown in \figref{clover:binary}.
\vspace{-2em}
\begin{figure}[h]
\centering    
\begin{lstlisting}[language=SQL, numbers=none]
SELECT * FROM R, S, T
WHERE R.x = S.x AND R.x = T.x AND S.x = T.x
\end{lstlisting}
\end{figure}
\vspace{-2em}


\subsection{\gj}\label{sec:background:gj}

The \gj algorithm, first introduced in~\cite{skew}, represents the simplest worst-case optimal join algorithm.
It builds upon the earlier Leapfrog Triejoin algorithm~\cite{leapfrog}.
The \gj algorithm computes the query $Q$ from \equref{conjq} by executing a series of nested loops, where each loop iterates over a specific query variable.
In particular, \gj arbitrarily selects a variable $x$, computes the intersection of all $x$-columns across the relations that contain $x$, and for each value $\theta$ in this intersection, it evaluates the residual query $Q[x/\theta]$.
In the residual query, every relation $R$ containing $x$ is replaced by \relx{R}{\theta} (or equivalently $R[\theta]$).
If the query $Q$ contains $k$ variables, the algorithm proceeds with $k$ nested loops.
In the innermost loop, \gj outputs a tuple consisting of constants derived from each iteration.


The \gj algorithm is provably worst-case optimal, achieving a time complexity of $O(n^{1.5})$, where $n$ represents the maximum possible size of the output~\cite{skew, agmbound}.
In contrast, binary join algorithms can exhibit a time complexity of $O(n^2)$.
While binary joins rely on hash tables for efficiency, \gj leverages a hash trie structure for each relation in the query.
A hash trie is a tree structure where each node is either a leaf node or a hash map that associates each atomic attribute value with another node.


\input{figures/clover_binary}

\subsection{\fj}\label{sec:background:fj}

A \fj plan defines how the \fj algorithm is executed, serving as a generalization and unification of both binary join plans and \gj plans~\cite{freejoin}.
In a left-deep linear plan for binary joins, the execution order is represented as a sequence of relations, where join attributes are implicitly determined by the shared attributes between relations.
In contrast, a \gj plan outlines a sequence of variables and does not explicitly reference the relations, as joins are performed on any relation sharing a particular variable.
A \fj plan, however, allows for the joining of any number of variables and relations at each step, requiring both to be explicitly specified.

Formally, a \fj plan is represented as a list $[\phi_1, \dotsc, \phi_m]$, where each $\phi_k$ is a list of subatoms from the query $Q$, referred to as a node.
The nodes must partition the query in the sense that for every atom \atom{R}{i}{x}{i} in the query, the set of all subatoms in all nodes must constitute a partitioning of \atom{R}{i}{x}{i}.
A subatom of an atom \atom{R}{i}{x}{i} is of the form \atom{R}{i}{y}{{}}, where $y$ is a subset of the variables $x_i$.
A partitioning of the atom \atom{R}{i}{x}{i} consists of subatoms $R_i(y_1), R_i(y_2), \dotsc$, where the sets $y_1, y_2, \dotsc$ form a partition of $x_i$.

To construct a \fj plan, the system begins with an optimized binary join plan produced by a traditional cost-based optimizer, such as DuckDB~\cite{duckdb, duckdb2, duckdb3}.
It first decomposes a bushy plan into a set of left-deep plans.
Each left-deep plan is then converted into an equivalent \fj plan.
After conversion, further optimization yields a plan that can range from a left-deep plan to a \gj plan.

Consider a plan derived from a straightforward conversion of the binary join plan for the clover query \clover into a \fj plan, as shown in \equref{clover:plan:naive}.
To execute the first node, we iterate over each tuple $(x, a)$ in $R$ and use $x$ to probe into $S$.
For each successful probe, we proceed to the second node, iterating over each value $b$ in $S[x]$, and then using $x$ to probe into $T$.
\begin{equation}
[[R(x, a), S(x)], [S(b), T(x)]]
\label{eq:clover:plan:naive}
\end{equation}

The plan corresponding to the \gj plan for the clover query \clover is depicted in \equref{clover:plan:gj}.
In this plan, execution starts by intersecting the sets $R.x \cap S.x \cap T.x$.
For each $x$ in the intersection, the values of $a$ and $b$ are retrieved from $R$ and $S$, respectively, and their Cartesian product is computed.
Additionally, after optimizing the naive plan from \equref{clover:plan:naive}, the resulting optimized \fj plan for the clover query \clover is shown in \equref{clover:plan:fj}.
\begin{equation}
[[R(x), S(x), T(x)], [R(a)], [S(b)]]
\label{eq:clover:plan:gj}
\end{equation}
\begin{equation}
[[R(x, a), S(x), T(x)], [S(b)]]
\label{eq:clover:plan:fj}
\end{equation}


\subsection{SDQL}\label{sec:background:sdql}

We utilize SDQL, a statically typed language capable of expressing relational algebra with aggregations, and functional collections over data structures such as relations using semi-ring dictionaries.
SDQL can serve as an intermediate language for data analytics, enabling the translation of programs written in relational algebra-based languages into SDQL.

\figref{clover:binary} illustrates the corresponding \sdql program to execute \clover using traditional binary joins, alongside its equivalent generated \cpp code.
In this example, we first join relations $R$ and $S$ (lines 1-11), followed by a join with relation $T$ (lines 14-24).
In \sdql, we use the !let! keyword to declare variables, such as a dictionary in line 1.
The !sum! keyword enables iteration over dictionary entries, as demonstrated in line 2.
Conditional logic is expressed with !if!, and membership is checked using the $\in$ operator to verify if an element exists in a dictionary (line 8).
If an element exists, its associated value can be accessed using the $(\ldots)$ syntax, as shown in line 9.
We employ \code{std::tuple} to implement records (line 11).

In addition to these basic and predefined syntaxes, we extended \sdql to meet our requirements by incorporating support for various dictionary types and underlying data structures for dictionary representation.
The subsequent sections will explore these extensions in detail.

%% file: figures/clover_binary.tex
\begin{figure*}[t]

\begin{minipage}{0.46\textwidth}
\begin{lstlisting}[language=sdql]
  let S_ht = 
    sum(<i, _> <- range(S.size)) 
      {S.x(i) -> {i -> 1}} in

  let RS = 
    sum(<R_i, _> <- range(R.size))
      let x = R.x(R_i) in
      if (x $\in$ S_ht)
        let Sx = S_ht(x) in
        sum (S_i, _> <- Sx)
          {<x=x, a=R.a(R_i), b=S.b(S_i)> -> 1} in

  let T_ht = 
    sum(<i, _> <- range(T.size)) 
      {T.x(i) -> {i -> 1}} in

  sum(<RS_i, _> <- range(RS.size))
    let x = RS.x(RS_i) in
    if (x $\in$ T_ht)
      let Tx = T_ht(x) in
      sum (T_i, _> <- Tx)
        {<x=x, a=RS.a(RS_i), b=RS.b(RS_i)> -> 1}
$$
\end{lstlisting}
\subcaption{\sdql.}
\label{fig:clover:binary:sdql}
\end{minipage}
\begin{minipage}{0.53\textwidth}
\begin{lstlisting}[language=C++]
  HT<int, HT<int, bool>> S_ht;
  for (int i = 0; i < S.size; ++i)
    S_ht[S.x[i]][i] += 1;

  HT<tuple<int, int, int>, int> RS;
  for (int R_i = 0; R_i < R.size; ++R_i) {
    auto x = R.x[R_i];
    if (S_ht.contains(x)) {
      auto &Sx = S_ht.at(x);
      for (auto &[S_i, S_v] : Sx)
        res[{x, R.a[R_i], S.b[S_i]}] += 1;
  }}

  HT<int, HT<int, bool>> T_ht;
  for (int i = 0; i < T.size; ++i)
    T_ht[T.x[i]][i] += 1;

  HT<tuple<int, int, int>, int> RST;
  for (auto &[RS_i, RS_v] : RS) {
    auto x = get<0>(RS[RS_i]);
    if (T_ht.contains(x)) {
      auto &Tx = S_ht.at(x);
      for (auto &[T_i, T_v] : Tx)
        RST[{x, get<1>(RS[RS_i]), get<2>(RS[RS_i])}] += 1;
  }}
\end{lstlisting}
\subcaption{\cpp.}
\label{fig:clover:binary:cpp}
\end{minipage}

\caption{Implementation of \clover based on traditional binary joins in \sdql and \cpp.}
\label{fig:clover:binary}
\end{figure*}

%% file: sections/3_system.tex
\section{System}\label{sec:system}

\begin{figure}[t]
\centering
\includegraphics[width=0.9\columnwidth]{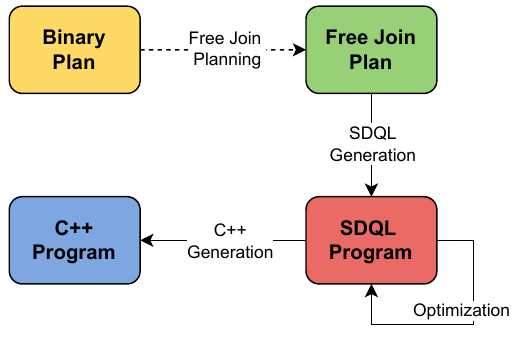}
\caption{An overview of our system architecture.}
\label{fig:arch}
\end{figure}

The methodology, as depicted in \figref{arch}, employs a multi-stage pipeline architecture.
The initial phase involves the transformation of a binary plan into a \fj plan (\secref{system:planning}).
Subsequently, we proceed to generate a naive \sdql program corresponding to the \fj plan (\secref{system:sdqlgen}).
Several optimizations are applied to enhance the performance of the initial SDQL program (\secref{opts}).
The final stage of the pipeline entails the generation of efficient \cpp code derived from the \sdql program, facilitating efficient query execution (\secref{system:cppgen}).

\subsection{Planning}\label{sec:system:planning}

This approach begins by taking an optimized binary join plan as input and transforming it into a \fj plan~\cite{freejoin}.
This transformation process is based on the methodology described in \secref{background:fj} and the \fj paper, which we utilize for consistency and fair comparison.
The system initially starts with an optimized binary plan and then converts it to an equivalent \fj plan.
The corresponding plans for the \gj and \fj algorithms for \clover would resemble those in \equref{clover:plan:gj} and \equref{clover:plan:fj}, respectively.


\input{figures/clover_gj}

\subsection{\sdql Program Generation}\label{sec:system:sdqlgen}

At this stage, we take the \fj plan from the previous step and generate an efficient \sdql program as an intermediate representation.
This process is divided into two phases: trie creation and query execution.

\smartpara{Trie Creation Phase} 
We construct the tries that will be utilized during the query execution phase.
Within \sdql, these tries function as nested hash maps, where each leaf node is a set of offsets into the base relation represented as a hash map with these offsets as the key and \code{true} as the value.
Each internal level is a hash map that maps attribute values to the next level’s hash map.

For instance, as illustrated in lines 2-10 of \figref{clover:gj:sdql} and \equref{clover:plan:gj}, we need a trie for each of the relations $R$, $S$, and $T$ in \clover, which align with the \gj algorithm.
These tries are created from attribute $x$ to the offsets of each relation, enabling access to other attributes in subsequent steps.
In contrast, the \fj implementation only requires building tries over relations $S$ and $T$ (lines 2-7 of \figref{clover:fj:sdql}) since we iterate directly over relation $R$.

\smartpara{Query Execution Phase} 
We generate an \sdql program that corresponds to the converted \fj plan, as described in \secref{system:planning}, utilizing the previously constructed tries.
The execution of each node in the \fj plan involves iterating over the first relation or its trie, depending on the plan, and using the attribute values to probe into each of the other tries.

In the example, for the first node in the plan for \gj $[R(x), S(x), T(x)]$, we iterate over $R$ and use $x$ values to probe into $S$ and $T$.
We do a similar process for the first node of the \fj's plan $[R(x, a), S(x), T(x)]$.
For each $x$ value successfully probed in both, we proceed to execute the second node.
This process is reflected in lines 14-17 and 11-15, corresponding to the first node of the plan in \figsref{clover:gj:sdql}{clover:fj:sdql}, respectively.
The subsequent nodes represent the Cartesian product among the other attribute values to make the final results for a given $x$ value. The translation of the nodes $[[R(a)], [S(b)]]$ are shown in lines 18-20 in \figref{clover:gj:sdql}, and the translation of $[S(b)]$ is shown in lines 16-17 in \figref{clover:fj:sdql}.

\input{figures/clover_fj}


\subsection{\cpp Code Generation}\label{sec:system:cppgen}

The final component of our pipeline involves generating \cpp code for \sdql programs, which is relatively straightforward, as illustrated in \figsref{clover:gj}{clover:fj}.
To enhance performance, summations that yield dictionaries are translated into loops that perform in-place updates.
In addition to the transformations outlined in \secref{background:sdql}, the subsequent sections will demonstrate the translation of the extended data structures.

%% file: figures/clover_gj.tex
\begin{figure*}[t]
\begin{minipage}[t]{\columnwidth}
\begin{lstlisting}[language=sdql]
   // Trie Creation
   let R_trie = 
     sum(<i, _> <- range(R.size)) 
       {R.x(i) -> {i -> 1}} in
   let S_trie = 
     sum(<i, _> <- range(S.size)) 
       {S.x(i) -> {i -> 1}} in
   let T_trie = 
     sum(<i, _> <- range(T.size)) 
       {T.x(i) -> {i -> 1}} in

   // Query Execution

   sum(<x, Rx> in R_trie)
     if (x $\in$ S_trie && x $\in$ T_trie) then
       let Sx = S_trie(x) in
       let Tx = T_trie(x) in
       sum(<R_i, _> <- Rx)
         sum(<S_i, _> <- Sx)
           {<c0=x, c1=R.a(R_i), c2=S.b(S_i)> -> 1}
$$
\end{lstlisting}
\subcaption{\sdql.}
\label{fig:clover:gj:sdql}
\end{minipage}
\begin{minipage}[t]{\columnwidth}
\begin{lstlisting}[language=C++]
   // Trie Creation
   HT<int, HT<int, bool>> R_trie;
   for (int i = 0; i < R.size; ++i)
     R_trie[R.x[i]][i] += 1;
   HT<int, HT<int, bool>> S_trie;
   for (int i = 0; i < S.size; ++i)
     S_trie[S.x[i]][i] += 1;
   HT<int, HT<int, bool>> T_trie;
   for (int i = 0; i < T.size; ++i)
     T_trie[T.x[i]][i] += 1;

   // Query Execution
   HT<tuple<int, int, int>, int> res;
   for (auto &[x, Rx] : R_trie)
     if (S_trie.contains(x) && T_trie.contains(x)) {
       auto &Sx = S_trie.at(x);
       auto &Tx = T_trie.at(x);
       for (auto &[R_i, R_v] : Rx)
         for (auto &[S_i, S_v] : Sx)
           res[{x, R.a[R_i], S.b[S_i]}] += 1;
     }
\end{lstlisting}
\subcaption{\cpp.}
\label{fig:clover:gj:cpp}
\end{minipage}
\caption{\gj implementation of \clover in \sdql and \cpp.}
\label{fig:clover:gj}
\end{figure*}

%% file: figures/clover_fj.tex
\begin{figure*}[t]
\begin{minipage}{\columnwidth}
\begin{lstlisting}[language=sdql]
   // Trie Creation
   let S_trie = 
     sum(<i, _> <- range(S.size)) 
       {S.x(i) -> {i -> 1}} in
   let T_trie = 
     sum(<i, _> <- range(T.size)) 
       {T.x(i) -> {i -> 1}} in

   // Query Execution

   sum(<R_i, _> in range(R.size)) 
     let x = R.x(R_i) in
     if (x $\in$ S_trie && x $\in$ T_trie) then
       let Sx = S_trie(x) in
       let Tx = T_trie(x) in
       sum(<S_i, _> <- Sx)
         {<c0=x, c1=R.a(R_i), c2=S.b(S_i)> -> 1}
$$
\end{lstlisting}
\subcaption{\sdql.}
\label{fig:clover:fj:sdql}
\end{minipage}
\begin{minipage}{\columnwidth}
\begin{lstlisting}[language=C++]
   // Trie Creation
   HT<int, HT<int, bool>> S_trie;
   for (int i = 0; i < S.size; ++i)
     S_trie[S.x[i]][i] += 1;
   HT<int, HT<int, bool>> T_trie;
   for (int i = 0; i < T.size; ++i)
     T_trie[T.x[i]][i] += 1;

   // Query Execution
   HT<tuple<int, int, int>, int> res;
   for (int R_i = 0; R_i < R.size; ++R_i) {
     auto x = R.x[R_i];
     if (S_trie.contains(x) && T_trie.contains(x)) {
       auto &Sx = S_trie.at(x);
       auto &Tx = T_trie.at(x);
       for (auto &[S_i, S_v] : Sx)
         res[{x, R.a[R_i], S.b[S_i]}] += 1;
   }}
\end{lstlisting}
\subcaption{\cpp.}
\label{fig:clover:fj:cpp}
\end{minipage}

\caption{\fj implementation of \clover in \sdql and \cpp.}
\label{fig:clover:fj}
\end{figure*}

%% file: sections/4_efficiency.tex
\section{Efficiency} \label{sec:opts}

In this section, we sequentially apply a series of optimizations to the naive implementation of \clover, building upon each previous one for better comprehension.
\secref{opts:dictspec} introduces dictionary specialization, optimizing the underlying data structures for dictionary representation.
\secref{opts:earlyproj} discusses the early projection techniques employed to enhance performance.
Finally, \secref{opts:sorting} presents our novel hybrid approach, alongside the support of sort-based WCOJ algorithms, to leverage the strengths of both hash-based and sort-based paradigms.


\subsection{Dictionary Specialization}\label{sec:opts:dictspec}

Dictionary specialization is a technique aimed at optimizing the data structures used to represent leaf nodes in tries, significantly improving both trie creation and query execution performance.
When a dictionary is created in SDQL, it serves two primary purposes: lookup and iteration.
When lookups are required, an underlying dictionary data structure is necessary.
However, if the operation only involves iterating over the dictionary, we can store only the keys in a more efficient data structure.
This subsection focuses on two specific optimizations we employed to enhance dictionary specialization in our system: \code{std::vector} and \smallvec.

\subsubsection{\textbf{Vector (O1)}}
Replacing hash maps with vectors for leaf nodes in tries can significantly improve performance by reducing the overhead of key-value pair operations.
As outlined in \secref{system:sdqlgen}, each leaf node in the tries was initially represented as a hash map that mapped offsets in the base relation to a constant boolean value (\code{true}).
However, since only the keys of these hash maps are required, as demonstrated in \figsref{clover:gj:sdql}{clover:fj:sdql}, this structure can be optimized by converting the key-value pair representation into a list implementation that stores only the keys.

This optimization is applied in SDQL programs through the use of !@vec! annotation to specify the dictionary representation, as shown in \figref{opt:vec:sdql}.
When this annotation is applied to a hash map, we employ the \code{VecDict} data structure, as shown in \figref{vecdict}, in \cpp, replacing the hash table that maps offsets to \code{true}, as illustrated in \figref{opt:vec:cpp}.
This data structure acts as a wrapper around \code{std::vector}, providing a dictionary-like interface.
Using a vector under the hood enhances performance by reducing the cost of both insertion and iteration, as operations in \code{std::vector} are generally less expensive than in hash tables.

\input{figures/vecdict}

\input{figures/vec_opt}

\subsubsection{\textbf{SmallVector (O2)}}
\smallvec is a specialized data structure designed to function as a vector-like container optimized for small sequences.
It improves performance by allocating storage on the stack, thus reducing the overhead of heap allocations and enhancing cache locality.
This optimization is particularly advantageous when the vector contains only a small number of elements.
\smallvec allocates a fixed number of elements on the stack, and when the number of elements exceeds this predefined size, it switches to heap allocation.
This strategy offers a balance between performance and flexibility, and implementations of \smallvec are used in systems such as Rust~\cite{rust_smallvec} and LLVM~\cite{llvm_smallvector}.

We implemented a custom version of \smallvec, as shown in \figref{smallvec} in \cpp to replace the underlying data structure of the leaf nodes used to store offsets.
As previously discussed, when performing lookups on relation $R$ for a given $\theta$ value, $R[\theta]$ typically contains a small number of elements.
In such cases, \smallvec enhances performance by avoiding heap allocation for nodes with few elements during trie creation.
Another application of this data structure occurs when building a trie on a unique attribute, such as primary keys, which is common in join operations.
Since this attribute is unique, each value appears only once, resulting in a single offset per value.
While a single variable could be used in this scenario, \smallvec effectively handles it.

The interface of \smallvec closely mirrors that of \code{std::vector}, allowing it to be seamlessly integrated in contexts where \code{std::vector} is typically used, as shown in \figref{opt:smallvec}.
\smallvec manages dynamic memory allocation internally, abstracting the complexity from the user while delivering improved efficiency.

\input{figures/smallvec}

\input{figures/smallvec_opt}


\subsection{Early Projection/Aggregation}\label{sec:opts:earlyproj}
Early projection and aggregation are techniques that reduce the amount of data processed and stored during query execution by identifying and eliminating unnecessary columns as early as possible.
This approach can significantly improve query performance by reducing memory usage, enhancing cache utilization, and accelerating join operations.
In this subsection, we discuss three specific optimizations that we employed in our system.

\subsubsection{\textbf{Dead Code Elimination (O3)}}
Dead code elimination is a powerful optimization technique that can significantly improve query performance by reducing the amount of data processed and stored during join operations.
This technique systematically removes unnecessary columns during the join process, whether these columns originate from base relations or intermediate results.
By identifying and eliminating attributes that are not required for subsequent operations or the final query output, the volume of data processed and stored throughout the query execution pipeline is minimized.
This reduction accelerates join operations by decreasing memory usage and enhancing cache utilization, leading to overall improvements in query efficiency.

\subsubsection{\textbf{Eliminating Redundant Offsets (O4)}}
For relations that do not contribute to the final query output, it is unnecessary to store their offsets during trie construction, as we do not need to access other attributes from these relations.
These relations are only utilized for joining and checking the existence of values for the attributes involved in the joins.
Therefore, eliminating redundant offsets can reduce the overhead associated with trie construction.

In the clover query \clover, for example, relation $T$ is used to join on attribute $x$ with relations $R$ and $S$, but it does not contribute any attributes to the final results.
For such relations, the primary task is to verify the existence of $x$ values from $R$, without accessing $T$'s attributes.
Therefore, storing offsets for a relation like $T$ becomes unnecessary.
To address this, we can optimize the hash map by replacing the value type from a vector-like data structure with an integer variable, as illustrated in \figref{opt:bool}.
This modification reduces the overhead associated with appending elements for relations involved solely in joins, thereby streamlining the trie creation process and enhancing overall efficiency.

\input{figures/bool_opt}

\subsubsection{\textbf{Loop-Invariant Code Motion (O5)}}
By identifying and moving invariant expressions out of loops, loop-invariant code motion can significantly reduce the number of operations required for aggregation calculations.

In the context of calculating the left-hand side of \equref{sumab}, we can observe that for each $\beta_j$, $\alpha_i$ is a constant value and can be moved outside the inner summation loop.
Similarly, now for each $\alpha_i$, the summation of all values in $\beta$ is also a constant value, allowing it to be moved outside the outer summation loop.
This optimization reduces the number of required multiplications from $n \times k$ to $1$ and the number of summations from $n \times k$ to $n + k$, resulting in a significant performance improvement.
\begin{equation}
\sum_{i=1}^n\sum_{j=1}^k (\alpha_i \times \beta_j) = \sum_{i=1}^n (\alpha_i \times (\sum_{j=1}^k \beta_j)) = (\sum_{i=1}^n \alpha_i) \times (\sum_{j=1}^k \beta_j)
\label{eq:sumab}
\end{equation}

As discussed in \secref{background}, projections and aggregations are not explicitly represented in \fj plans.
For instance, consider an aggregation on query \clover, which projects only the minimum values of attributes $a$ and $b$.
The naive implementation of these aggregations is shown in the deleted lines (with light red background) of \figref{opt:licm}.
In this approach, for each $x$ value that satisfies the join conditions, $2 \times k$ \code{min} operations are performed, assuming the size of $S$'s offsets is $k$.
By applying loop-invariant code motion, we can move the minimum operation for attribute $a$ outside the loop over $S$'s offsets, reducing the number of minimum operations by $k$.
This leaves only $k$ operations to find the minimum value of $b$, plus two additional operations to update the final output.
This optimization effectively reduces unnecessary operations and improves the overall efficiency of the aggregation process.

\input{figures/licm_opt}


\subsection{Sorting vs Hashing}\label{sec:opts:sorting}

The realm of worst-case optimal joins is characterized by two primary paradigms: hash-based approaches, exemplified by Umbra~\cite{umbra} and \fj~\cite{freejoin}, and sort-based approaches, such as Leapfrog Triejoin~\cite{leapfrog}, EmptyHeaded~\cite{emptyheaded}, and LMFAO~\cite{lmfao}.
Our system is designed to efficiently support both paradigms, allowing for the execution of sort-based \wcoj algorithms alongside hash-based methods.
For the sort-based approach, we assume that input relations are always provided in sorted order.

\figref{clover:sorting} illustrates the implementation of the sort-based approach for query \clover in both \sdql and \cpp.
In our system, the !@st! annotation is used to specify that the dictionary is a sorted dictionary.
For the \cpp implementation, we developed a custom sorted dictionary data structure, as shown in \figref{sorteddict}.
Since we assume that the input relations are sorted by the attributes involved in the joins, insertions always occur at the end, or we update the last elements in the sorted dictionary.
For lookups, we employ binary search to efficiently locate a given key among the sorted keys and return its corresponding value.

\input{figures/sorteddict}

Another optimization we employed in this case is the use of the !@range! annotation.
When a relation is sorted by an attribute $x$, all occurrences of each $x$ value appear consecutively.
Instead of storing all occurrence offsets for each $x$ value in a vector-like structure, we optimize by only keeping the first and last offsets of this consecutive block of elements, as shown in in \figref{range}.
To insert an offset into this structure, we simply update the right bound of the range.
During iteration, we can efficiently loop from the left bound to the right bound, reducing both storage overhead and iteration complexity.

\input{figures/range}

As discussed earlier, we decompose a bushy plan into a set of left-deep plans, and each left-deep plan produces an intermediate result.
There is no guarantee that these intermediate results will remain sorted.
For such cases, we must first sort the intermediate results before using them in trie creation with our sorted dictionary data structure.
To reduce the overhead of sorting, we employ a hybrid approach by using a hash table for each intermediate result, bypassing the need for sorting them.
While binary search in a sorted dictionary is less efficient than lookups in hash tables, the sorted dictionary proves advantageous during the trie creation phase, which is often the time-consuming part of a query.
This allows the trie creation phase for intermediate results to remain more efficient, even when utilizing sorted dictionaries for base relations.

\input{figures/clover_sorting}

%% file: figures/vecdict.tex
\begin{figure}[t]
\begin{minipage}[t]{\columnwidth}
\begin{lstlisting}[language=C++]
   class VecDict {
     vector<T> vec; // SmallVector in SmallVecDict
     class Proxy {
       VecDict &vecdict;
       T key;
       void operator+=(int) {
         vecdict.vec.push_back(key); 
     }};
     Proxy operator[](T key) {
       return Proxy(*this, key);
   }};
\end{lstlisting}
\end{minipage}
\caption{\code{VecDict} data structure.}
\label{fig:vecdict}
\end{figure}

%% file: figures/vec_opt.tex
\begin{figure*}[t]
\begin{minipage}{\columnwidth}
\begin{lstlisting}[language=sdql]
   // Trie Creation
   let S_trie = 
     sum(<i, _> <- range(S.size)) 
\end{lstlisting}
\begin{lstlisting}[language=sdql, backgroundcolor=\color{deleted}, firstnumber=4]
 -   S.x(i) -> {i -> 1}} in
\end{lstlisting}
\begin{lstlisting}[language=sdql, backgroundcolor=\color{added}, firstnumber=4]
 +   {S.x(i) -> @vec {i -> 1}} in
\end{lstlisting}
\begin{lstlisting}[language=sdql, firstnumber=5]
   let T_trie = 
     sum(<i, _> <- range(T.size)) 
\end{lstlisting}
\begin{lstlisting}[language=sdql, backgroundcolor=\color{deleted}, firstnumber=7]
 -   {T.x(i) -> {i -> 1}} in
\end{lstlisting}
\begin{lstlisting}[language=sdql, backgroundcolor=\color{added}, firstnumber=7]
 +   {T.x(i) -> @vec {i -> 1}} in
\end{lstlisting}
\begin{lstlisting}[language=sdql, firstnumber=8]

   // Query Execution

   sum(<R_i, _> in range(R.size)) 
     let x = R.x(R_i) in
     if (x $\in$ S_trie && x $\in$ T_trie) then
       let Sx = S_trie(x) in
       let Tx = T_trie(x) in
       sum(<S_i, _> <- Sx)
         {<c0=x, c1=R.a(R_i), c2=S.b(S_i)> -> 1}
$$
\end{lstlisting}
\subcaption{\sdql.}
\label{fig:opt:vec:sdql}
\end{minipage}
\begin{minipage}{\columnwidth}
\begin{lstlisting}[language=C++]
   // Trie Creation
\end{lstlisting}
\begin{lstlisting}[language=C++, backgroundcolor=\color{deleted}, firstnumber=2]
 - HT<int, HT<int, bool>> S_trie;
\end{lstlisting}
\begin{lstlisting}[language=C++, backgroundcolor=\color{added}, firstnumber=2]
 + HT<int, VecDict<int>> S_trie;
\end{lstlisting}
\begin{lstlisting}[language=C++, firstnumber=3]
   for (int i = 0; i < S.size; ++i)
     S_trie[S.x[i]][i] += 1;
\end{lstlisting}
\begin{lstlisting}[language=C++, backgroundcolor=\color{deleted}, firstnumber=5]
 - HT<int, HT<int, bool>> T_trie;
\end{lstlisting}
\begin{lstlisting}[language=C++, backgroundcolor=\color{added}, firstnumber=5]
 + HT<int, VecDict<int>> T_trie;
\end{lstlisting}
\begin{lstlisting}[language=C++, firstnumber=6]
   for (int i = 0; i < T.size; ++i)
     T_trie[T.x[i]][i] += 1;

   // Query Execution
   HT<tuple<int, int, int>, int> res;
   for (int R_i = 0; R_i < R.size; ++R_i) {
     auto x = R.x[R_i];
     if (S_trie.contains(x) && T_trie.contains(x)){
       auto &Sx = S_trie.at(x);
       auto &Tx = T_trie.at(x);
\end{lstlisting}
\begin{lstlisting}[language=C++, backgroundcolor=\color{deleted}, firstnumber=16]
 -     for (auto &[S_i, S_v] : Sx)
\end{lstlisting}
\begin{lstlisting}[language=C++, backgroundcolor=\color{added}, firstnumber=16]
 +     for (auto &S_i : Sx)
\end{lstlisting}
\begin{lstlisting}[language=C++, firstnumber=17]
         res[{x, R.a[R_i], S.b[S_i]}] += 1;
   }}
\end{lstlisting}
\subcaption{\cpp.}
\label{fig:opt:vec:cpp}
\end{minipage}

\caption{Impact of using Vector data structure for inner dictionaries which store offsets into base relations.}
\label{fig:opt:vec}
\end{figure*}

%% file: figures/smallvec.tex
\begin{figure}[t]
\begin{minipage}[t]{\columnwidth}
\begin{lstlisting}[language=C++]
   class SmallVector {
     array<T, N> stack;
     vector<T> *heap;
     size_t size{0};
     void push_back(const T &value) {
       if (size++ < N)
         stack[size] = value;
       else {
         if (size++ == N) {
           heap = new vector<T>(
             stack.begin(), stack.end());
         }
         heap->push_back(value);
   }}};
\end{lstlisting}
\end{minipage}
\caption{\code{SmallVector} data structure.}
\label{fig:smallvec}
\end{figure}

%% file: figures/smallvec_opt.tex
\begin{figure*}[t]
\begin{minipage}{\columnwidth}
\begin{lstlisting}[language=sdql]
   // Trie Creation
   let S_trie =
     sum(<i, _> <- range(S.size)) 
\end{lstlisting}
\begin{lstlisting}[language=sdql, backgroundcolor=\color{deleted}, firstnumber=4]
 -   {S.x(i) -> @vec {i -> 1}} in
\end{lstlisting}
\begin{lstlisting}[language=sdql, backgroundcolor=\color{added}, firstnumber=4]
 +   {S.x(i) -> @smallvec(4) {i -> 1}} in
\end{lstlisting}
\begin{lstlisting}[language=sdql, firstnumber=5]
   let T_trie =
     sum(<i, _> <- range(T.size)) 
\end{lstlisting}
\begin{lstlisting}[language=sdql, backgroundcolor=\color{deleted}, firstnumber=7]
 -   {T.x(i) -> @vec {i -> 1}} in
\end{lstlisting}
\begin{lstlisting}[language=sdql, backgroundcolor=\color{added}, firstnumber=7]
 +   {T.x(i) -> @smallvec(4) {i -> 1}} in
\end{lstlisting}

\subcaption{\sdql.}
\label{fig:opt:smallvec:sdql}
\end{minipage}
\begin{minipage}{\columnwidth}
\begin{lstlisting}[language=C++]
   // Trie Creation
\end{lstlisting}
\begin{lstlisting}[language=C++, backgroundcolor=\color{deleted}, firstnumber=2]
 - HT<int, VecDict<int>> S_trie;
\end{lstlisting}
\begin{lstlisting}[language=C++, backgroundcolor=\color{added}, firstnumber=2]
 + HT<int, SmallVecDict<int, 4>> S_trie;
\end{lstlisting}
\begin{lstlisting}[language=C++, firstnumber=3]
   for (int i = 0; i < S.size; ++i)
     S_trie[S.x[i]][i] += 1;
\end{lstlisting}
\begin{lstlisting}[language=C++, backgroundcolor=\color{deleted}, firstnumber=5]
 - HT<int, VecDict<int>> T_trie;
\end{lstlisting}
\begin{lstlisting}[language=C++, backgroundcolor=\color{added}, firstnumber=5]
 + HT<int, SmallVecDict<int, 4>> T_trie;
\end{lstlisting}
\begin{lstlisting}[language=C++, firstnumber=6]
   for (int i = 0; i < T.size; ++i)
     T_trie[T.x[i]][i] += 1;
\end{lstlisting}
\subcaption{\cpp.}
\label{fig:opt:smallvec:cpp}
\end{minipage}

\caption{Impact of using SmallVector data structure for inner dictionaries which store offsets.}
\label{fig:opt:smallvec}
\end{figure*}

%% file: figures/bool_opt.tex
\begin{figure*}[t]
\begin{minipage}{\columnwidth}
\begin{lstlisting}[language=sdql]
   // Trie Creation
   let S_trie = 
     sum(<i, _> <- range(S.size))
       {S.x(i) -> @smallvec(4) {i -> 1}} in
   let T_trie = 
     sum(<i, _> <- range(T.size)) 
\end{lstlisting}
\begin{lstlisting}[language=sdql, backgroundcolor=\color{deleted}, firstnumber=7]
 -   {T.x(i) -> @smallvec(4) {i -> 1}} in
\end{lstlisting}
\begin{lstlisting}[language=sdql, backgroundcolor=\color{added}, firstnumber=7]
 +   {T.x(i) -> 1} in
\end{lstlisting}


\subcaption{\sdql.}
\label{fig:opt:bool:sdql}
\end{minipage}
\begin{minipage}{\columnwidth}
\begin{lstlisting}[language=C++]
   // Trie Creation
   HT<int, SmallVecDict<int, 4>> S_trie;
   for (int i = 0; i < S.size; ++i)
     S_trie[S.x[i]][i] += 1;
\end{lstlisting}
\begin{lstlisting}[language=C++, backgroundcolor=\color{deleted}, firstnumber=5]
 - HT<int, SmallVecDict<int, 4>> T_trie;
\end{lstlisting}
\begin{lstlisting}[language=C++, backgroundcolor=\color{added}, firstnumber=5]
 + HT<int, int> T_trie;
\end{lstlisting}
\begin{lstlisting}[language=C++, firstnumber=6]
   for (int i = 0; i < T.size; ++i)
\end{lstlisting}
\begin{lstlisting}[language=C++, backgroundcolor=\color{deleted}, firstnumber=7]
 -   T_trie[T.x[i]][i] += 1;
\end{lstlisting}
\begin{lstlisting}[language=C++, backgroundcolor=\color{added}, firstnumber=7]
 +   T_trie[T.x[i]] += 1;
\end{lstlisting}

\subcaption{\cpp.}
\label{fig:opt:bool:cpp}
\end{minipage}

\caption{Impact of redundant offsets elimination for join-only relations.}
\label{fig:opt:bool}
\end{figure*}

%% file: figures/licm_opt.tex
\begin{figure*}[t]
\begin{minipage}{\columnwidth}


\begin{lstlisting}[language=sdql, firstnumber=9]
   // Query Execution

   sum(<R_i, _> in range(R.size)) 
     let x = R.x(R_i) in
     if (x $\in$ S_trie && x $\in$ T_trie) then
       let Sx = S_trie(x) in
       let Tx = T_trie(x) in
\end{lstlisting}
\begin{lstlisting}[language=sdql, backgroundcolor=\color{deleted}, firstnumber=16]
 -     sum(<S_i, _> <- Sx)
 -       promote[min_sum](<c0=R.a(R_i),c1=S.b(S_i)>)
\end{lstlisting}
\begin{lstlisting}[language=sdql, backgroundcolor=\color{added}, firstnumber=16]
 +     let R_mn = <c0=R.a(R_i)> in
 +     let S_mn = 
 +       sum(<S_i, _> <- Sx)
 +         promote[min_sum](<c0=S.b(S_i)>) in
 +     promote[min_sum](<c0=R_mn.c0, c1=S_mn.c0>)
\end{lstlisting}
\begin{lstlisting}[language=sdql, firstnumber=21]
$$
\end{lstlisting}
\subcaption{\sdql.}
\label{fig:opt:licm:sdql}
\end{minipage}
\begin{minipage}{\columnwidth}
\begin{lstlisting}[language=sdql, firstnumber=9]
   // Query Execution
\end{lstlisting}
\begin{lstlisting}[language=C++, backgroundcolor=\color{deleted}, firstnumber=10]
 - HT<tuple<int, int, int>, int> res;
\end{lstlisting}
\begin{lstlisting}[language=C++, backgroundcolor=\color{added}, firstnumber=10]
 + tuple<int, int> res {INF, INF};
\end{lstlisting}
\begin{lstlisting}[language=C++, firstnumber=11]
   for (int R_i = 0; R_i < R.size; ++R_i) {
     auto x = R.x[R_i];
     if (S_trie.contains(x) && T_trie.contains(x)) {
       auto &Sx = S_trie.at(x);
       auto &Tx = T_trie.at(x);
\end{lstlisting}
\begin{lstlisting}[language=C++, backgroundcolor=\color{deleted}, firstnumber=16]
 -     for (auto &S_i : Sx)
 -       min_inplace(res, {R.a[R_i], S.b[S_i]});
\end{lstlisting}
\begin{lstlisting}[language=C++, backgroundcolor=\color{added}, firstnumber=16]
 +     tuple<int> R_mn {R.a[R_i]};
 +     tuple<int> S_mn {INF};
 +     for (auto &S_i : Sx)
 +       min_inplace(S_mn, {S.b[S_i]});
 +     min_inplace(res,{get<0>(R_mn),get<0>(S_mn)});
\end{lstlisting}
\begin{lstlisting}[language=C++, firstnumber=21]
   }}
\end{lstlisting}
\subcaption{\cpp.}
\label{fig:opt:licm:cpp}
\end{minipage}

\caption{Impact of loop-invariant code motion on aggregation operations.}
\label{fig:opt:licm}
\end{figure*}

%% file: figures/sorteddict.tex
\begin{figure}[t]
\begin{minipage}[t]{\columnwidth}
\begin{lstlisting}[language=C++]
   class SortedDict {
     // stores keys and values in std::vector
     // values of type VT are int or Range
     vector::iterator find(const KT &key) {
       /* std::lower_bound binary search */
   }};
\end{lstlisting}
\end{minipage}
\caption{\code{SortedDict} data structure.}
\label{fig:sorteddict}
\end{figure}

%% file: figures/range.tex
\begin{figure}[t]
\begin{minipage}[t]{\columnwidth}
\begin{lstlisting}[language=C++]
   class Range {
     size_t left;
     size_t right;
     class Proxy {
       Range &range;
       void operator+=(int) { ++range.right; }
     };
     Proxy operator[](size_t const idx) {
       if ( /* is first access */ )
         left = right = idx;
       return Proxy(*this);
   }};
\end{lstlisting}
\end{minipage}
\caption{\code{Range} data structure.}
\label{fig:range}
\end{figure}

%% file: figures/clover_sorting.tex
\begin{figure*}[t]

\begin{minipage}{\columnwidth}
\begin{lstlisting}[language=sdql]
   // Trie Creation
   let S_trie =
     sum(<i, _> <- range(S.size))
       @st {S.x(i) -> @range {i -> 1}} in
   let T_trie = 
     sum(<i, _> <- range(T.size))
       @st {T.x(i) -> @range {i -> 1}} in
\end{lstlisting}
\subcaption{\sdql.}
\label{fig:clover:sorting:sdql}
\end{minipage}
\begin{minipage}{\columnwidth}
\begin{lstlisting}[language=C++]
   // Trie Creation
   SortedDict<int, Range> S_trie;
   for (int i = 0; i < S.size; ++i)
     S_trie[S.x[i]][i] += 1;
   SortedDict<int, Range> T_trie;
   for (int i = 0; i < T.size; ++i)
     T_trie[T.x[i]][i] += 1;
\end{lstlisting}
\subcaption{\cpp.}
\label{fig:clover:sorting:cpp}
\end{minipage}

\caption{Sort-based implementation of the trie creation phase of \clover in \sdql and \cpp.}
\label{fig:clover:sorting}
\end{figure*}

%% file: sections/5_experiments.tex
\section{Experiments}\label{sec:exp}

We implemented our system in a three-step pipeline.
First, we take a binary join plan, produced and optimized by DuckDB, converting it into a \fj plan~\cite{freejoin}.
Then, we translate the \fj plan into an \sdql program, which serves as our intermediate representation, and apply various optimizations.
Finally, we generate C++ code from the optimized \sdql program to execute the query.

We compare our approach against the \fj framework~\cite{freejoin} on both \gj and \fj implementations, recognizing it as the state-of-the-art system that outperforms in-memory databases such as DuckDB~\cite{duckdb, duckdb2, duckdb3}. 
For an apple-to-apple comparison, we use the same query plans as the \fj framework~\cite{freejoin}.
To evaluate performance, we use the widely adopted Join Order Benchmark (JOB)~\cite{job} and the LSQB benchmark~\cite{lsqb}.
Three research questions guide our evaluation:
\begin{enumerate}
    \item How does our system compare to the \gj and \fj implementations of \fj framework? (\secref{exp:runtime})
    \item What is the impact of optimizations we employed? (\secref{exp:opts})
    \item How do the hash-based and sort-based approaches perform in our system? (\secref{exp:sorting})
\end{enumerate}


\subsection{Setup}

Both the JOB and LSQB benchmarks are primarily focused on evaluating join performance.
The JOB benchmark consists of 113 acyclic queries, with an average of 8 joins per query, while the LSQB benchmark includes a mix of cyclic and acyclic queries.
Each query in both benchmarks involves base-table filters, natural joins, and a simple group-by operation at the end.
JOB operates on real-world data from the IMDB dataset, whereas LSQB uses synthetic data.
For a fair comparison, we executed all benchmarks on the query set reported by the \fj framework, which serves as our competitor. 
The only exception is query Q3 from the LSQB benchmark, which we excluded as the results are not reproducible using the \fj framework's open-source implementation.

All experiments were conducted on a MacBook Pro running macOS 15.0.1, equipped with an Apple M1 Max chip and 64GB of LPDDR5 RAM.
Each experiment was executed 5 times and the average run times were reported.
All systems were configured to run in single-threaded mode and operate entirely in main memory.
We employed an efficient hash table implementation in C++ known as phmap~\cite{phmap}.
All code was compiled using Clang 18.1.8 with the following flags:
\begin{lstlisting}[numbers=none]
-std=c++17 -O3 -march=native -mtune=native -Wno-narrowing -ftree-vectorize
\end{lstlisting}


\subsection{Performance Comparison}\label{sec:exp:runtime}

Our first set of experiments compares the performance of our system for \gj and \fj algorithms against the \fj framework on both JOB and LSQB benchmarks.

\subsubsection{\textbf{JOB}}\label{sec:exp:runtime:job}

\figref{exp:job} presents a run time comparison of our system with the \fj framework, evaluating both the \gj and \fj algorithms on JOB queries.
Since our system does not currently support vectorization, \figref{exp:sorting:hybrid:wo} illustrates the performance of our system relative to the non-vectorized version of the \fj framework, which employs the same underlying algorithm.

In \figref{exp:job}, the majority of data points for \gj and both non-vectorized and vectorized versions of \fj algorithms appear below the diagonal, indicating that our system outperforms the \fj framework in these cases.
This suggests that, despite the lack of vectorization support, our system outperforms the performance of the \fj framework.

On average (geometric mean), our system demonstrates a speedup of $1.49\times$ and $1.42\times$ over the \fj framework for the \gj and \fj algorithms, respectively, and achieves a $2.70\times$ performance improvement over the non-vectorized version of \fj.
The maximum speedups observed are $3.14\times$ for \gj and $4.78\times$ for \fj, while the minimum speedups are $0.71\times$ ($40\%$ slowdown) and $0.30\times$ ($3.33\times$ slowdown), respectively.



As discussed in \secref{system:sdqlgen}, our system requires that tries be fully constructed before query execution begins, meaning the data structure we currently employ does not support lazy evaluation, unlike \fj~\cite{freejoin}.
However, as outlined in \secref{system:planning}, we utilize the same execution plans produced and used by the \fj framework.
In these plans, when a relation appears as the first relation in a node, we iterate over its offsets to access its attribute values.
At this stage, all attribute values for that relation are made available to subsequent nodes in the execution plan, thereby eliminating the need for further iterations or lookups.

Once we identify the first node where each relation is used for iteration, we construct a trie with levels corresponding to the attributes of that relation that appeared in earlier nodes, since lookups for those attribute values are required.
For all queries in the JOB benchmark, each relation involves at most one attribute lookup before iteration.
This means that each relation is either used for iteration or accessed first for lookups over a single attribute, followed by iteration over the offsets linked to that attribute.
Consequently, our approach behaves similarly to leveraging lazy data structures, as the construction of the first level of tries for non-iterated relations is necessary, and lookups over these relations are guaranteed.

\input{figures/job_results}

\subsubsection{\textbf{LSQB}}\label{sec:exp:runtime:lsqb}

\figref{exp:lsqb} presents a performance comparison between our system and the \fj framework for both \gj and \fj algorithms on LSQB queries.
Each line in the figure represents a query executed across scaling factors of 0.1, 0.3, 1, and 3.
It is important to note that the \fj framework encountered an error when running Q3, and we were unable to reproduce its results for this query.

For Q2, which is a cyclic query, our system outperforms the \fj framework across scaling factors, achieving speedups of up to $2.49\times$ (on average $2.15\times$) for the \gj algorithm and up to $1.50\times$ (on average $1.28\times$) for the \fj algorithm.
Contrary to our discussion about JOB queries in \secref{exp:runtime:job}, there is a relation in Q2 that necessitates a trie with a depth greater than one.
While our approach to trie construction in this scenario is less efficient than using lazy data structures, our system still demonstrates superior performance, with a significant performance gap compared to the \fj framework.

For the acyclic queries, our system’s performance is comparable to the \fj framework for Q4.
In the case of Q5, our system is up to $1.60\times$ (on average $1.49\times$) faster for the \gj algorithm.
However, for the \fj algorithm, the performance remains similar, with our system being slightly faster at smaller scaling factors and slightly slower at larger ones.

A significant performance improvement is observed for Q1, where our system achieves speedups of up to $27.52\times$ (on average $8.50\times$) for \gj and $80.05\times$ (on average $23.13\times$) for \fj compared to the \fj framework.
We investigated the $80\times$ speedup, which occurs for scaling factor 0.3, and found it to show up repeatedly across benchmark runs.
While we recognize the potential for a $10\times$ speedup through factorization in the \fj framework, we were unable to reproduce their results.
Even if we assume the \fj framework achieves this speedup, our system would still maintain a considerable performance advantage based on the aforementioned speedups.

Overall, this substantial gap is primarily attributed to the early projection and aggregation optimizations integrated into our system.
Unlike the JOB queries, in LSQB, the output size before aggregation is significantly larger than the input size, resulting in a considerable amount of time spent on output construction.
Our system mitigates this overhead by pushing projection and aggregation earlier in the query execution process.

\input{figures/lsqb_results}


\subsection{Impact of Optimisations}\label{sec:exp:opts}

\begin{figure}[t]
    \centering
    \includegraphics[width=0.95\columnwidth]{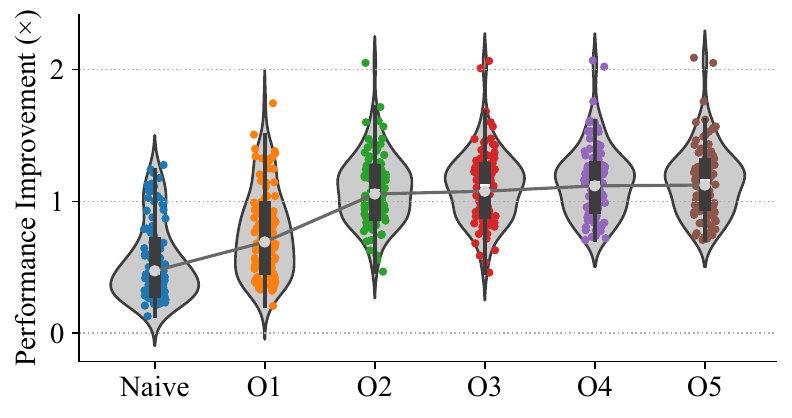}
    \caption{Impact of optimizations. Each point shows the performance improvement of a query over the \fj framework. Each violin is the distribution of the performance improvements for all queries after applying the given optimization. The gray line shows the geometric mean for each optimization.}
    \label{fig:exp:violin}
\end{figure}

As discussed in \secref{opts}, we implemented a series of optimizations to enhance the efficiency of our naive implementation.
\figref{exp:violin} illustrates the cumulative effect of these optimizations, showing the distribution of performance improvements relative to the \fj framework for JOB queries.
Initially, without any optimizations, our naive implementation of the \fj algorithm was $2.11\times$ slower than the \fj framework.
Each subsequent optimization progressively narrowed this gap, contributing to the overall performance gains observed in our system.

After applying O1, we observe an improvement in performance, though our system remains slower than the \fj framework.
With the application of O2, our system, despite lacking support for lazy data structures and vectorization, slightly outperforms the \fj framework with a $1.056\times$ speedup.
Up to this point, the impact of each optimization is clearly visible in \figref{exp:violin}, and these optimizations are also available in the \fj framework.

In the violin plot for O3, the lower part of the distribution (below the median) becomes thinner, while the upper part thickens, indicating a shift toward better performance.
Additionally, we observe two data points with more than $2\times$ speedup rather than one in the previous optimization, and our overall speedup increases to $1.077\times$.
With the application of O4, the tail of the O3 distribution is eliminated, resulting in an increased average speedup of $1.117\times$.

O5 further improves the performance of all queries slightly compared to O4.
Ultimately, our fully optimized implementation achieves a $1.124\times$ speedup, which is $2.38\times$ faster than the naive implementation and $6.5\%$ faster than O2.
The subtle speedups in O3, O4, and O5 are attributed to the fact that trie construction dominates the overall run time for most of the queries.
However, these optimizations are built on top of earlier optimizations targeting trie creation and only focus on improving the query execution phase.

\begin{figure}[t]
    \centering
    \includegraphics[width=0.95\columnwidth]{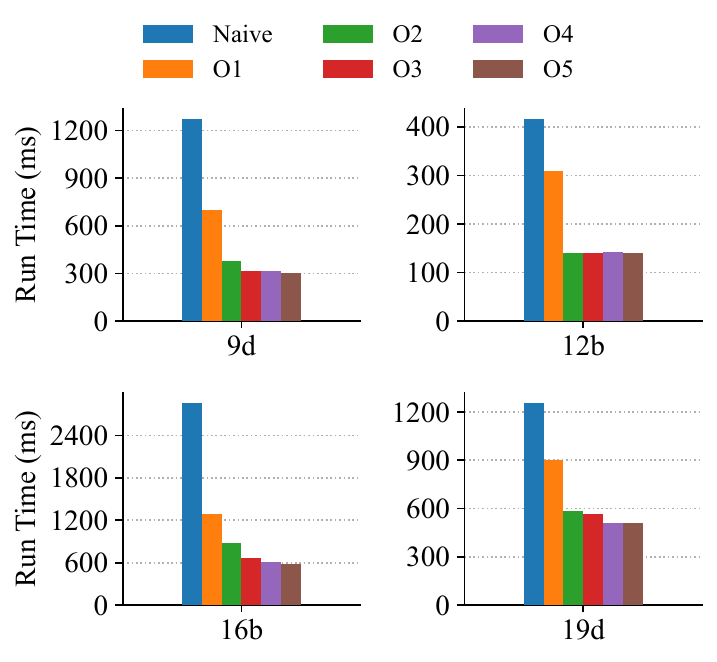}
    \caption{Ablation study. Each bar shows the run time of a query in our system after applying its corresponding optimization. O1: \code{std::vector}. O2: SmallVector. O3: Dead Code Elimination. O4: Eliminating Redundant Offsets. O5: Loop-Invariant Code Motion.}
    \label{fig:exp:ablation}
\end{figure}

We analyzed the impact of optimizations on a representative subset of queries, starting with our naive implementation and progressively applying each optimization.
The results, shown in \figref{exp:ablation}, demonstrate that all optimizations introduced in \secref{opts} contribute positively to their respective scenarios.
The O1 and O2 optimizations highlight the benefits of using \code{std::vector} and SmallVector, which enhance the performance of all selected queries.
The O3 optimization adds the Dead Code Elimination, affecting queries 9d and 16b.
The O4 optimization further improves the previous ones by eliminating redundant offsets, which affects queries 16b and 19d.
Finally, in O5, we apply Loop-Invariant Code Motion, which improves the performance of queries 9d and 16b.

\subsection{Hash-based vs Sort-based Performance}\label{sec:exp:sorting}

In \figref{exp:job:fj}, illustrates the performance of the hash-based approach in our system.
Data points, which represent the comparison between our hash-based implementation and \fj, cluster around the diagonal.
This suggests that the hash-based approach of our system matches and slightly outperforms the performance of the \fj framework.
Specifically, our system achieves a better performance up to $2.09\times$ (on average $1.12\times$) than the \fj framework.

As discussed in \secref{opts:sorting}, our system also supports the sort-based paradigm of worst-case optimal join (WCOJ) algorithms.
For this class of algorithms, we assume that the input data is always provided in sorted order.
\figref{exp:sorting:pure} presents the performance of our sort-based implementation for all JOB queries in comparison to the \fj framework.
Our sort-based approach demonstrates performance improvements of up to $6.25\times$ (on average $1.07\times$).
As can be realized, these two approaches are algorithmically distinct, which explains why the data points are not concentrated around the diagonal in \figref{exp:sorting:pure}.


As mentioned earlier, even when input data is sorted, there is no guarantee that intermediate results will remain sorted.
In such cases, we must sort these intermediate results before constructing their trie using a sorted dictionary, which can lead to significant overhead in the overall execution time.
To address this, we introduce a novel hybrid approach that utilizes sorted dictionaries for base relations that are already sorted, while using hash tables for intermediate results.
This eliminates the need to sort intermediate results.
Using this hybrid approach, we achieve superior performance over the \fj framework for almost all queries in the JOB benchmark, as shown in \figref{exp:sorting:hybrid}.
Specifically, our hybrid approach demonstrates a performance improvement of up to $4.78\times$ (on average $1.42\times$) compared to \fj.

\input{figures/sorting_results}

\begin{figure}[t]
    \centering
    \includegraphics[width=0.95\columnwidth]{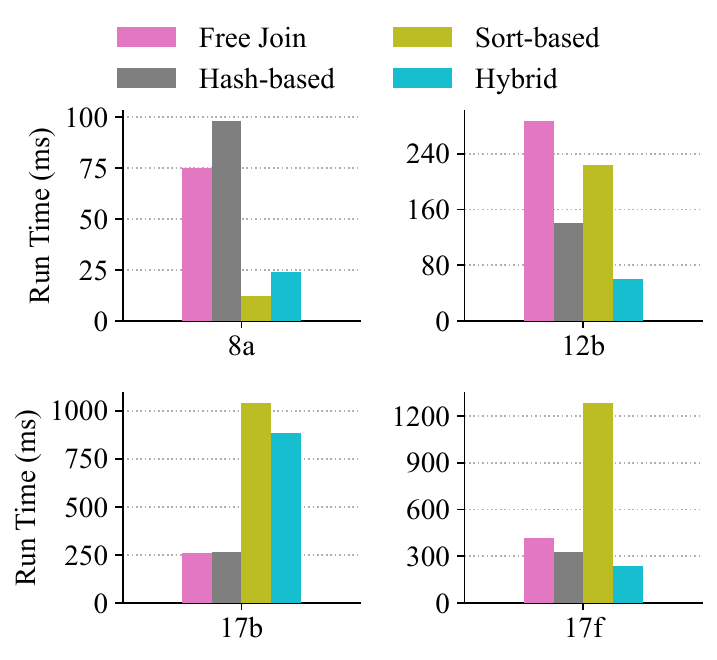}
    \caption{Run time comparison among \fj and the hash-based, sort-based, and hybrid implementations in our system. Each bar shows the performance of an alternative on the given query.}
    \label{fig:exp:alters}
\end{figure}

For most queries in the JOB benchmark, trie creation is the most time-consuming aspect when using hash tables.
By employing sorted dictionaries via \code{SortedDict} and \code{Range} data structures--where only the first and last offsets of an element's occurrences are stored instead of a vector-like structure--we can significantly improve run time performance.
This approach reduces the overhead associated with hash tables, such as allocation, insertions, and updates.
However, the use of binary search for lookups in sorted dictionaries can slow down query execution for cases where a significant portion of the run time is spent on the query execution phase itself, as seen in the queries above the diagonal in \figref{exp:sorting:hybrid}.
The efficiency of each approach—hash-based or sort-based—depends on both the input data and the specific query being executed.
However, our system provides the flexibility to utilize any of these approaches, enabling the efficient execution of any query on any dataset.

\figref{exp:alters} highlights the run time of a representative subset of queries, providing deeper insights into the scenarios where each approach--hash-based, sort-based, or hybrid--performs better.
For instance, query 12b, one of the points above the $2\times$ line in \figref{exp:violin}, benefits from the hybrid approach, resulting in the highest speedups among all JOB queries.
This improvement is largely due to the efficient handling of intermediate results using hash tables, which avoids the overhead of sorting.
However, when dealing with smaller relations (after applying filters), sorting is relatively fast.
In such cases, the hybrid approach's advantages may not fully offset the overhead introduced by hash tables.
Query 8a is an example of this; as shown in \figref{exp:alters}, the sort-based solution offers the best performance due to the small size of the relations, making sorting more efficient.

Queries 17b and 17f provide an interesting comparison, as they share the same joins but apply different filters to their base relations.
Both queries slow down significantly with the sort-based approach, largely because they involve a large number of lookups where binary search negates the speedup gained from trie creation.
However, the hybrid approach manages to compensate for the performance in query 17f but not in 17b.
This disparity stems from the order in which lookups are applied.
In query 17b, the plan probes a base relation first, followed by an intermediate result.
Even with a hash table for the intermediate result, binary search is still used for all elements.
In contrast, query 17f first probes the intermediate result using the hash table, where lookups are performed in a constant time.
Only for the elements that find a match in the intermediate result does the query then perform binary searches on the base relation, reducing the number of $O(\log{n})$ lookups, resulting in a more efficient execution.

%% file: figures/job_results.tex
\begin{figure*}[t]
\begin{subfigure}{0.30\textwidth}
\includegraphics[width=\textwidth]{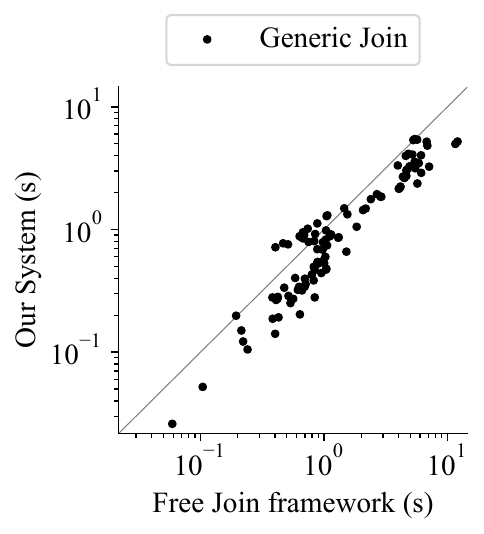}
\caption{\gj.}
\label{fig:exp:job:gj}
\end{subfigure}
\hspace{1em}
\begin{subfigure}{0.30\textwidth}
\includegraphics[width=\textwidth]{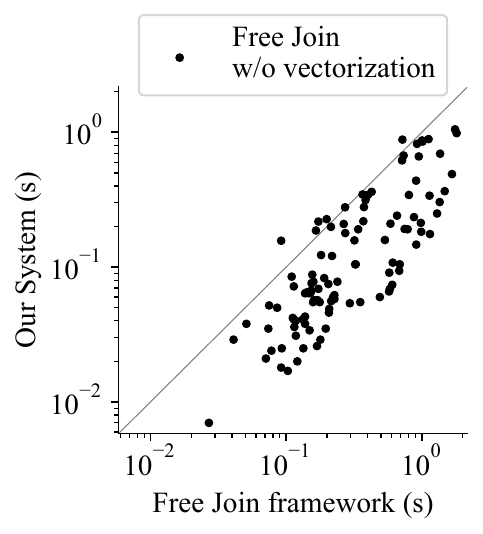}
\caption{\fj without vectorization.}
\label{fig:exp:sorting:hybrid:wo}
\end{subfigure}
\hspace{1em}
\begin{subfigure}{0.30\textwidth}
\includegraphics[width=\textwidth]{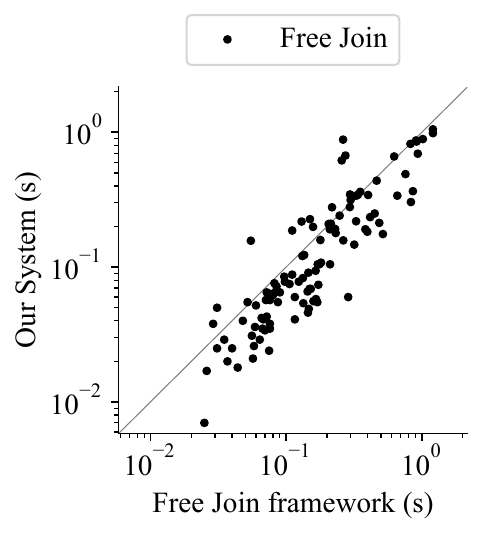}
\caption{\fj.}
\label{fig:exp:sorting:hybrid}
\end{subfigure}

\caption{Run-time comparison on JOB. Each point compares the run time of a query on our system and \fj framework. \figref{exp:job:gj} compares the \gj implementation of each system. \figsref{exp:sorting:hybrid:wo}{exp:sorting:hybrid} compare our system's \fj implementation with the \fj framework~\cite{freejoin} without and with vectorization. Each point below the diagonal line represents a query for which our system is faster.}
\label{fig:exp:job}
\end{figure*}

%% file: figures/lsqb_results.tex
\begin{figure}[t]
\begin{subfigure}{0.47\columnwidth}
\includegraphics[width=\textwidth]{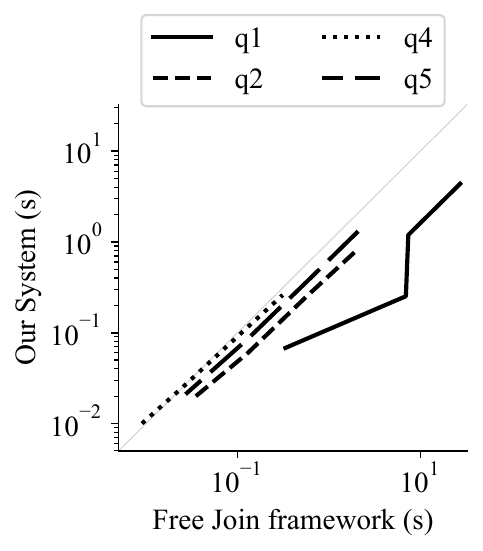}
\caption{\gj.}
\label{fig:exp:lsqb:gj}
\end{subfigure}
\hspace{1em}
\begin{subfigure}{0.47\columnwidth}
\includegraphics[width=\textwidth]{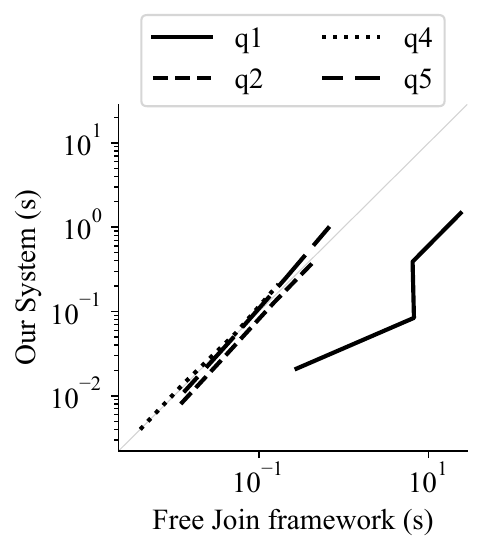}
\caption{\fj.}
\label{fig:exp:lsqb:fj}
\end{subfigure}

\caption{Runtime comparison on LSQB.  Each line is a query running on increasing scaling factors (0.1, 0.3, 1, 3) and compares our system and \fj framework. \figref{exp:lsqb:gj} compares the \gj implementation of each system. \figref{exp:lsqb:fj} compares our system's \fj implementation with \fj framework.}
\label{fig:exp:lsqb}
\end{figure}

%% file: figures/sorting_results.tex
\begin{figure}[t]
\begin{subfigure}{0.47\columnwidth}
\includegraphics[width=\textwidth]{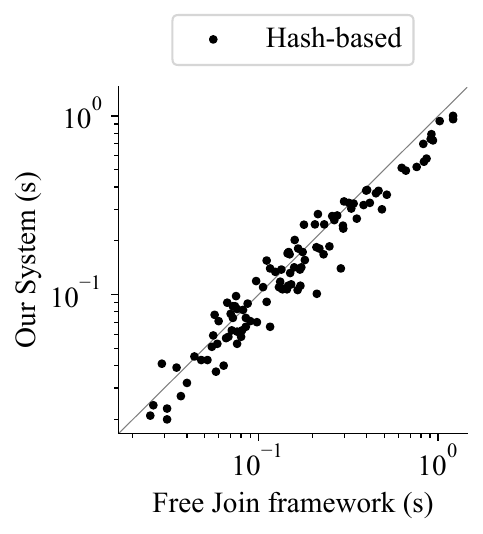}
\caption{Hash-based.}
\label{fig:exp:job:fj}
\end{subfigure}
\hspace{1em}
\begin{subfigure}{0.47\columnwidth}
\includegraphics[width=\textwidth]{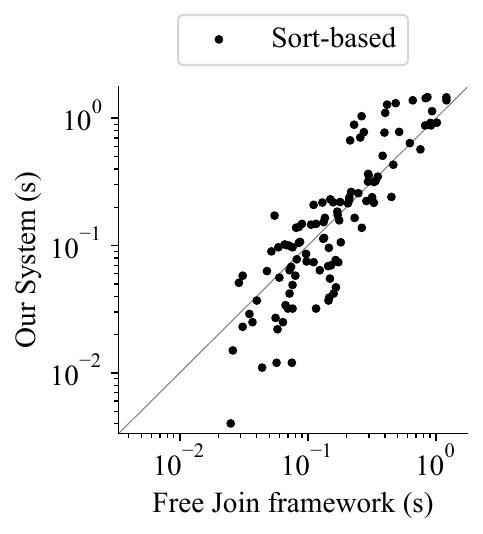}
\caption{Sort-based.}
\label{fig:exp:sorting:pure}
\end{subfigure}

\caption{Run-time comparison between sort- and hash-based approaches and \fj on JOB. \figref{exp:job:fj} compares the performance of the hash-based approach implemented in our system. \figref{exp:sorting:pure} compares the performance of the sort-based approach.}
\label{fig:exp:sorting}
\end{figure}

%% file: sections/6_conclusion.tex
\section{Conclusion and Future Work}\label{sec:futurework}

In this paper, we introduce a unified architecture that integrates binary join and worst-case optimal join (\wcoj) algorithms, as well as hash-based and sort-based \wcoj methods.
Our proposed system consistently outperforms or matches state-of-the-art solutions across all hash-based, sort-based, and hybrid approaches.
This flexibility allows for selecting the most efficient method tailored to the input data and the executing query.

For future research, we envision four primary directions.
First, our system has limitations compared to the state-of-the-art, particularly \fj~\cite{freejoin}, as previously discussed.
Enhancing performance through support for lazy data structures and vectorization is one area for further improvement.
The second direction involves adding parallelism to our system.
For instance, the query execution phase could be processed in chunks, as everything, except the final results, is read-only, requiring only the merging of final results.
Similarly, parallelism could be exploited during the trie creation phase in the sort-based approach, since merging sorted dictionaries is straightforward.

We currently use binary search for lookups in sorted dictionaries, which has a time complexity of $O(\log{n})$, asymptotically less efficient than the corresponding operation in hash tables.
Another area of improvement is implementing hinted lookups~\cite{hinted} for sorted dictionaries, which would allow us to achieve the same amortized time complexity as hash tables when the input data used for lookups is also already sorted.

Currently, our system utilizes the query optimizer developed by DuckDB and \fj.
However, this optimizer is not ideally suited for our system, as existing optimizers are typically designed to focus exclusively on either hash-based or sort-based approaches.
An integrated optimizer tailored to our architecture could potentially generate more efficient query plans based on the specific algorithm employed.
A notable example of this can be seen in the performance comparison between queries 17b and 17f.
While the hash-based approach optimizes the order of lookups solely based on the relation sizes, the hybrid approach introduces an additional parameter that must be considered to produce an optimal execution plan.